\documentclass[aip,apl,amsmath,amssymb,reprint,twocolumn]{revtex4-2}

\usepackage{amsmath,amssymb,amsfonts,amsthm}
\usepackage{graphicx,xcolor,transparent}
\usepackage{physics}
\usepackage{tikz}
\usetikzlibrary{calc,positioning}

\newcommand{\C}{\mathbb{C}}
\newcommand{\R}{\mathbb{R}}
\newcommand{\Z}{\mathbb{Z}}
\newcommand{\N}{\mathbb{N}}

\newcommand{\OO}{\mathcal{O}}

\newcommand{\PT}{\ensuremath{\mathcal{PT}}}
\newcommand{\T}{\ensuremath{\mathcal{T}}}

\newcommand{\bi}[1][i]{\mathbf{t}_{#1}}
\newcommand{\ri}[1][i]{\mathbf{g}_{#1}}

\newcommand{\ba}{\mathbf{t}_1}
\newcommand{\bb}{\mathbf{t}_2}
\newcommand{\bk}{\mathbf{k}}
\newcommand{\br}{\mathbf{r}}

\newcommand{\ra}{\mathbf{g}_1}
\newcommand{\rb}{\mathbf{g}_2}

\usepackage[colorlinks,linkcolor=red,citecolor=blue]{hyperref}

\begin{document}

\title{Nodal phases in non-Hermitian wallpaper crystals}

\author{J. Lukas K. König}
\email[]{lukas.konig@fysik.su.se}
\author{Felix Herber}
\email[]{fehe3416@student.su.se}
\author{Emil J. Bergholtz}
\email[]{emil.bergholtz@fysik.su.se}
\thanks{author to whom correspondence should be addressed}
\affiliation{Department of Physics, Stockholm University, AlbaNova University Center, 106 91 Stockholm, Sweden}

\date{\today}

\begin{abstract}
    Symmetry and non-Hermiticity play pivotal roles in photonic lattices. While symmetries such as parity-time ($\mathcal{PT}$) symmetry have attracted ample attention, more intricate crystalline symmetries have been neglected in comparison. Here, we investigate the impact of the 17 wallpaper space groups of two-dimensional crystals on non-Hermitian band structures. We show that the non-trivial space group representations enforce degeneracies at high symmetry points and dictate their dispersion away from these points. In combination with either $\mathcal T$ or $\mathcal{PT}$, the symmorphic p4mm symmetry and the non-symmorphic p2mg, p2gg, and p4gm symmetries protect exceptional chains intersecting at the pertinent high symmetry points. 
\end{abstract}

\maketitle

\begin{figure*}
    \centering
    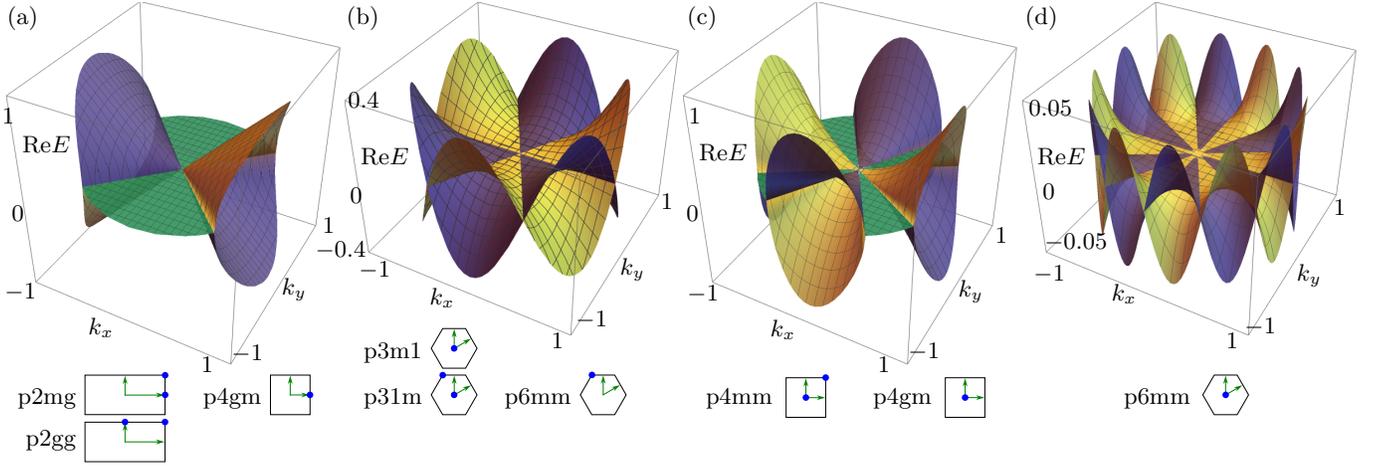
    \caption{
    Main result: 
    Non-trivial point group representation for the indicated wallpaper groups enforces nodal structure around points \(\bk^\star\) of high symmetry.
    Under \PT{} symmetry, it further enforces nodal lines emanating from \(\bk^\star\).
    (a) Typical local low-energy dispersion around points with \(D_2^\star\) point group obtained from Eq.~\eqref{eq:space-group-d2-only} for \(E_1=E_2=1\) features two lines of exceptional points.
    Inset: corresponding Brillouin zones, with points \(\bk^\star\) demarcated in blue.
    (b)-(d) as (a), for points with a \(D_3\) (\(D_4\), \(D_6\)) point group; 
    models given in Eqs.~\eqref{eqs:pt-models} with \(A=\tilde B_1= \tilde B_2 = C = 1\). 
    Three nodal (four exceptional, six nodal) lines cross at \(\bk^\star\).
    }
    \label{fig:main-result}
\end{figure*}

In the past decades, non-Hermitian effective descriptions have emerged as a central framework used to describe a wide variety of experimental setups \cite{ashidaNonHermitianPhysics2020}.
Generally, the non-Hermitian terms describe gain or dissipation in the respective systems. 
The applications can range from classical metamaterials through photonic crystals and quantum optics to open quantum systems. 

The spectra of such systems differ significantly from their Hermitian counterparts \cite{bergholtzExceptionalTopologyNonHermitian2021}.
For one, non-Hermitian operators feature degenerate points with higher abundance; their codimension is two compared to three for Hermitian degeneracies \cite{berryPhysicsNonhermitianDegeneracies2004,heissPhysicsExceptionalPoints2012}. As such non-Hermitian degenerate points occur generically in two-dimensions.
Furthermore, these degenerate points are  deficient, so-called exceptional points (EPs), and feature non-differentiable dispersion relations \cite{katoPerturbationTheoryLinear1995,zhouObservationBulkFermi2018,berryPhysicsNonhermitianDegeneracies2004,heissPhysicsExceptionalPoints2012,koziiNonHermitianTopologicalTheory2017}.
In three dimensions they occur as exceptional lines that may form knots, links, and similar structures of topological interest \cite{carlstromExceptionalLinksTwisted2018, carlstromKnottedNonHermitianMetals2019, cerjanExperimentalRealizationWeyl2019, xuWeylExceptionalRings2017, yangJonesPolynomialKnot2020, zhangTidalSurfaceStates2021}.
EPs occur readily in photonic systems, more so if a perfect balance of gain and loss renders these systems \PT-symmetric \cite{miriExceptionalPointsOptics2019,regensburgerParityTimeSynthetic2012,zhangMultipleExceptionalPoints2023}.

In general, imposing symmetries \cite{
chiuClassificationTopologicalQuantum2016,ruiNonHermitianSpatialSymmetries2022} on a given system increases the abundance and variety of stable nodal points. In non-Hermitian systems, symmetries such as \PT{}  lead to EPs of codimension one, i.e. occurring generically in one dimension, forming stable lines in two dimensions and so on 
\cite{budichSymmetryprotectedNodalPhases2019, kawabataClassificationExceptionalPoints2019, kimuraChiralsymmetryProtectedExceptional2019, okugawaTopologicalExceptionalSurfaces2019, szameitMathcalMathcalSymmetry2011, yoshidaSymmetryprotectedExceptionalRings2019, zhouExceptionalSurfacesPTsymmetric2019,yangHomotopySymmetryNonHermitian2023,dingNonHermitianTopologyExceptionalpoint2022}. Moreover, it increases the number of degenerate energy levels which may lead to the generic appearance of higher-order EPs \cite{mandalSymmetryHigherOrderExceptional2021,delplaceSymmetryProtectedMultifoldExceptional2021,sayyadRealizingExceptionalPoints2022,stalhammarClassificationExceptionalNodal2021,huNonHermitianSwallowtailCatastrophe2023,crippaFourthorderExceptionalPoints2021,wangExperimentalSimulationSymmetryprotected2023}.
Considering internal symmetries, the Hermitian Altland-Zirnbauer classification must be extended to a 38-fold symmetry classification for non-Hermitian systems \cite{kawabataSymmetryTopologyNonHermitian2019,zhouPeriodicTableTopological2019}.
In Hermitian systems, it is well-known that spatial symmetries \cite{fuTopologicalCrystallineInsulators2011} can also enhance the abundance of nodal points, 
constrain them to particular shapes such as nodal chains \cite{bzdusekNodalchainMetals2016},
or enforce degenerate points at material surfaces \cite{wiederWallpaperFermionsNonsymmorphic2018}, 
or at specific points of high symmetry
\cite{bradlynDiracWeylFermions2016,kruthoffTopologicalClassificationCrystalline2017}.

Early results have extended these results to non-Hermitian systems, extending symmetry indicators \cite{shiozakiSymmetryIndicatorNonHermitian2021},
establishing the existence of crystalline-symmetry-protected nodal chains in three-dimensional systems \cite{zhangSymmetryprotectedTopologicalExceptional2023,cuiExperimentalRealizationStable2023}, and studying \PT -symmetric optical lattices using the representation theory of colored groups \cite{mockCharacterizationParitytimeSymmetry2016}
and the symmetry-protected degenerate points in two-dimensional square photonic lattices \cite{mockComprehensiveUnderstandingParitytime2017}. 
Since EPs appear generically already in two dimensions, we expect a general characterization of two-dimensional symmetry-enforced features to be of interest.

Here, we provide this characterization by listing all possible nodal structures that can be enforced by the two-dimensional space groups.
We find that seven of these groups feature points of high symmetry with non-trivial little group representations. 
These constrain the Hamiltonian to be degenerate at the points of high symmetry, featuring a Hermitian-type nodal point at which eigenvectors are orthogonal (in contrast to EPs where they are degenerate).
Taking into account time reversal symmetry \T, or a combination of time reversal and spatial inversion symmetries \PT, intrinsically non-Hermitian nodal structures emerge in the surrounding dispersion.
In particular, nodal points are enforced at the high-symmetry momenta of space group p4mm and p4gm (p2mg, p4gm, and p2gg) and turn out to be the intersection of four (two) exceptional lines.
They resemble non-defective EPs, seen as limit points along these exceptional lines \cite{sayyadSymmetryprotectedExceptionalNodal2023}.

We begin our work by recapitulating the physical intuition of space groups and how they give rise to and constrain the momentum-space Hamiltonian of crystalline systems. 
We continue by listing all space groups and momenta with non-trivial point group action, before explicitly enumerating the allowed Bloch Hamiltonians in the neighborhood of these points to lowest order.
We repeat this analysis for Bloch Hamiltonians constrained by spatial as well as either \PT{} or \T{} symmetry.
Finally, we present simple lattice models that illustrate our main results.

We describe our process and results within the tight-binding approximation:
states are expressed in a basis of orbitals at points in \(\R^2\). 
In crystalline systems these are arranged periodically in a lattice, i.e., there are two linearly independent shortest lattice vectors \(\ba, \bb\in\R^2\) such that for each state at \(\br\), there exist other states at \(\br+\Z\ba+\Z\bb\).
We, thus, label the orbitals by indices \((n,m)\in\Z^2\), plus another index \(j\) labeling the unit cell pattern that is repeated over the lattice.
Altogether, the Hilbert space consists of states \(\ket{\phi_{n,m,j}}\) that describe an orbital of the same type as \(\ket{\phi_{0,0,j}}\), shifted by \(n\ba+m\bb\).
The position associated with each basis state is then \(\br(n,m,j) = n\ba+m\bb+\br_j\) for some \(\br_j\in\R^2\).
We will denote the bases as \(\ket{\phi_j(\br)}\), suppressing the dependencies of \(\br\).

Lattice translations are represented by commuting operators \(T_{\ba}, T_{\bb}\) that permute orbitals,
\begin{equation}
    T_{\ba}^n T_{\bb}^m \ket{\phi_j(\br)} = \ket{\phi_j(\br+n\ba+m\bb)}.
    \label{eq:translation}
\end{equation}
Depending on the given crystal, there exist further spatial transformations with simple action on the given position basis. 
Geometrically, they correspond to isometries \(g\) on the Euclidean plane, and so are either rotations, reflections, or glide axes.
Their unitary action on Hilbert space consists of two parts: 
first, a state's position is changed by an operation \(u_g: \br(n,m,j)\mapsto \br'=\br(n',m',l)\), 
and second, the different states within the same unit cell may be mixed by a unitary \(U_g\).
For example, a reflection may exchange two orbitals.
The total action reads
\begin{equation}
    \mathcal{U}_g \ket{\phi_j(\br(n,m,j))} 
    = 
    (U_g)_{jl} \ket{\phi_l(\br(n',m',l))}.
\end{equation}
 
Together, translations and other isometries form the so-called \emph{wallpaper group}, or space group, of a given crystal.
There are seventeen of these groups in two-dimensional systems.

We will work in momentum space, i.e., transform to a basis of Hilbert space in which the translation operators are diagonal. 
To make this mathematically coherent, we work in a lattice of finite size under periodic boundary conditions
\(\br\sim \br+ N_a \ba \sim \br + N_b\bb\), before performing the thermodynamic limit \(N_a,N_b\to\infty\).
A Fourier transform of the position basis,
\begin{equation}
    \ket{\phi_j(\bk)} =
    \frac{1}{\sqrt{N_aN_b}} 
    \sum_{n,m} e^{-i \bk \cdot \br(n,m,j)} \ket{\phi_j(\br(n,m,j))},
    \label{eq:fourier-state}
\end{equation}
satisfies \(T_{\bi} \ket{\phi_j(\bk)} = e^{i \bk\cdot\bi} \ket{\phi_j(\bk)}\).
The dependence on crystalline momentum \(\bk\) is periodic with period \(\ra,\rb\), two so-called reciprocal lattice vectors .
These satisfy \(\ri \cdot \bi[j] = 2\pi\delta_{ij}\) and are given, e.g., in Table~III in Ref.~\onlinecite{cracknellTablesIrreducibleRepresentations1974}.
The momentum \(\bk\) takes values in \((2\pi/N_a \Z,2\pi/N_b \Z)\), which turns continuous in the thermodynamic limit.

A generic spatial transformation \(g\) acts on spatial coordinates as \( u_g :\br \mapsto \Delta_g \br + \delta_g\). Its point group part \(\Delta_g\) is an orthogonal matrix, and \(\delta_g\) is some shift, potentially a fraction of a lattice vector. 
On the momentum basis, \(g\) acts as
\begin{equation}
    \mathcal{U}_g \ket{\phi_j(\bk) } 
    = 
    e^{i \Delta_g \bk \cdot \delta_g} (U_g)_{jl} \ket{\phi_l(\Delta_g \bk)}.
    \label{eq:point-symmetry-on-state}
\end{equation}

Physically, we describe a crystal by its Hamiltonian operator, for which the spatial transformations that leave the crystal invariant must constitute a symmetry.
Translation invariance requires the single-particle Hamiltonian with matrix elements
\begin{equation}
    \mel{\phi_{n',m',j'}}{H}{\phi_{n,m,j}} = 
    H_{j'j}(\br(n',m',j'),\br(n,m,j))
\end{equation}
to have position dependence \(H_{j'j}(\br'-\br)\).
This block-diagonalizes the Hamiltonian in the momentum basis: 
by Eq.~\eqref{eq:fourier-state},  \( \mel{\psi_{j'}(k')}{H}{\psi_j(k)} = H_{j'j}(k) \delta(k-k') \) in the thermodynamic limit.
The blocks are continuous in \(\bk\), collectively called Bloch Hamiltonian, and have matrix elements
\begin{equation}
    H_{j'j}(k) = \sum_{\substack{ {\mathbf{\rho}=\br(n,m,j)-\br_{j'}} \\ n,m}} e^{-2i \bk\cdot \mathbf{\rho}} H_{j'j}(\mathbf{\rho}).
\end{equation}

Symmetry under non-translational spatial transformations, acting according to Eq.~\eqref{eq:point-symmetry-on-state}, constrains the Bloch Hamiltonian further by
\begin{equation}
    H(\bk) = U_g H(\Delta_g \bk) U_g^\dagger.
    \label{eq:symmetry-constraint-around}
\end{equation}
The space group constraints carry over directly to Bloch Hamiltonians if \(\Delta_g\) itself is a space group element.
This is not the case for so-called non-symmorphic groups. There, the occurring matrices \(U_g\) are not representations of the point group generated by elements \(\Delta_g\), and we treat these groups separately.

Altogether, the non-translation symmetries of a crystal effectively decrease the unit cell by determining the Bloch Hamiltonian at points \(H(\Delta\bk)\) given \(H(\bk)\). 
Equation~\eqref{eq:symmetry-constraint-around} constitutes a particularly strong constraint at points of high symmetry \(\bk^\star\) that satisfy \( \bk^\star = \Delta_g \bk^\star\) for some space group elements \(g\). 
At these points,
\begin{equation}
    H(\bk^\star) = U_g H(\bk^\star) U_g^\dagger,
    \label{eq:constrain-ham}
\end{equation}
and all group operations that keep \(\bk^\star\) invariant form its so-called little group.
We investigate this constraint in the following.

Note that the unitary matrices appearing in Eq.~\eqref{eq:constrain-ham} form a (projective) representation of the little group, as \(U_{gh} = U_{g} U_{h}\).
Such representations decompose into irreducible representations, which means that there exist bases in which \(H(\bk)\) is block diagonal and blockwise constrained by a simplest set of matrices \(U_g^{(i)}\).
We may, thus, focus on these blocks individually and assume the constraining unitary matrices in Eq.~\eqref{eq:constrain-ham} are irreducible representations of the little group.

\begin{table}[htbp]
    \centering
    \caption{Wallpaper groups \(G\) (crystallographic notation) with high-symmetry momenta \(\bk^\star\) (in multiples of reciprocal lattice vectors) that can feature non-trivial representations of their respective little group \(G^{\bk^\star}\) constraining Bloch Hamiltonian \(H(\bk^\star)\).
    The final column denotes whether the momenta \(\bk^\star\) are invariant under time reversal \T. 
    In p6mm, a rotation about \(2\pi/6\) maps the threefold rotation center at \(K\) to a different point that must thus carry the same representation. 
    These are the \(K, K'\) points known from graphene.
    The corresponding space group representations are given in Eqs.~\eqref{eqs:results-spatial-only}.
    }
    \begin{tabular}{cccc}
        \(G\) & \(\bk^\star\) & \(G^{\bk^\star}\) & \T-inv.
        \\
        \hline
        p3m1, p31m & 
            \(\Gamma=(0,0)\) &
            \(D_3\) &
            yes
        \\
        p31m, p6mm*&
            \(K=(-\frac13,\frac23)\)&
            \(D_3\) &
            no
        \\
        p4mm, p4gm &
            \(\Gamma=(0,0)\)&
            \(D_4\) &
            yes
        \\
        p4mm &
            \(M=(\frac12,\frac12)\)&
            \(D_4\) &
            yes
        \\
        p6mm &
            \(\Gamma=(0,0)\) &
            \(D_6\) &
            yes
        \\
        \hline
        p2mg &
            \(S =(\frac12,\frac12)\)&
            \(D_2^\star\)&
            yes
        \\
        p2gg &
            \(Y=(0,\frac12)\)&
            \(D_2^\star\)&
            yes
        \\
        p2mg, p2gg, p4gm & 
            \(X=(\frac12,0)\) &
            \(D_2^\star\) &
            yes
        \\
        p4gm &
            \(M=(\frac12,\frac12)\) &
            \(D_4\) &
            yes
    \end{tabular}
    \label{tab:high-sym-points}
\end{table}

We list all points of high symmetry that carry a non-trivial representation of the respective wallpaper group in Table~\ref{tab:high-sym-points}.
We note that there exist additional one-dimensional representations, which impose trivial constraints on \(H(\bk)\).
In the context of photonics, we further consider only bosonic representations, i.e., ones that map \(\ket{\phi(k)}\mapsto e^{2\pi i}\ket{\phi(k)}\) under a full rotation. 
A list including fermionic and one-dimensional representations is given in Ref.~\onlinecite{cracknellTablesIrreducibleRepresentations1974}.

We begin by discussing the symmorphic wallpaper groups.
The only little groups with non-trivial irreducible representations are the dihedral groups \(D_n, n\in\{3,4,6\}\). 
They are generated by a rotation \(r\) of order \(n\), a reflection \(m\), and relations
\begin{equation}
    D_n = \braket{r,m}{r^n,m^2,mrmr}.
\end{equation}
The non-trivial irreducible representations are two-dimensional and generated by
\begin{equation}
    \rho_h(r)=\operatorname{diag}(\omega,\omega^{-1})\qq{and}
    \rho_h(m)=\sigma_x
\end{equation}
where \(\omega=e^{2\pi i h / n}\), \(\sigma_x\) is the standard Pauli matrix, and \(0<h<\left\lceil{\frac n2}\right\rceil\in\N\) labels inequivalent representations\cite{bradleyMathematicalTheorySymmetry2010}.

All these representations constrain \(H(\bk^\star)\) in the same way, forcing it to be twofold degenerate:
applying Eq.~\eqref{eq:constrain-ham} with \(U_g\) chosen to be the rotation or reflection generator imposes \(H(\bk^\star)=d_0 I_2 + d_z\sigma_z\) and \(H(\bk^\star)=d_0 I_2 + d_x\sigma_x\), respectively.
In combination, these constraints require \(H(\bk^\star)=d_0 I_2\).
This means that Bloch Hamiltonians must be degenerate under this symmetry, \(E_1(\bk^\star)=E_2(\bk^\star)=d_0\), regardless of their Hermiticity. 

It is impossible to open a gap at $\bk^*$ by adding terms that have appropriate \(D_n\) symmetry to individual bands, since it is the orbitals' spatial structure that determines which representation they transform under. 
The remaining degree of freedom, \(d_0 I_2\), does not affect the gap structure.

For non-symmorphic wallpaper groups, this argument is complicated slightly.
These groups contain glide reflections, combinations of a reflection \(\Delta_g\) and a subsequent shift along the reflection axis.
Their point group part, the reflection \(\Delta_g\), is not itself a symmetry of the crystal.
It appears in Eq.~\eqref{eq:symmetry-constraint-around}, hence, while glide reflections have no fixed points in real space, the entire glide reflection axis constitutes fixed points in momentum space.
Correspondingly, intersections of such axes are points of higher symmetry.
The irreducible representations in terms of \(U_g\) for the relevant groups at these points come in two distinct classes and are listed explicitly in Ref.~\onlinecite{cracknellTablesIrreducibleRepresentations1974}.

One class of representations is given by
\begin{equation}
    \rho^\star(r)=\sigma_x,\qq{and}
    \rho^\star(g_x)=\sigma_y,
\end{equation}
where \(r\) is a rotation about \(\pi\) and \(g_x\) is a glide reflection along the \(x\)-axis, which occurs at points where two reflection axes intersect.
This is a representation of a non-Abelian double cover of \(D_2\) in which one of the reflection transformations squares to \(-1\). We denote this as \(D_2^\star\).
The second class occurs at the center of the Brillouin zone of \(p4gm\), where two reflection axes and two glide reflection axes intersect.
The representation is equivalent to the one of \(D_4\), except for a factor of \(-1\) as for the case of \(D_2^\star\), which is irrelevant in Eq.~\eqref{eq:symmetry-constraint-around}.
All of these representations enforce degenerate \(H(\bk^\star)\).

\begin{figure*}[htbp]
    \centering
    \def\countx{2} 
    \def\county{2} 
    \def\nodeasize{2em}
    \def\nodebsize{2em}
    \begin{tikzpicture}[
            sitea/.style = {circle, fill=none, draw=none, minimum size=\nodeasize, inner sep=0pt, outer sep=0pt},
            siteb/.style = {circle, fill=none, draw=none, minimum size=\nodebsize, inner sep=0pt, outer sep=0pt},
            orba/.style = {fill=orange,draw=none},
            orbb/.style = {fill=blue,draw=none},
            hopone/.style = {green!60!black,thick,->},
            hoptwo/.style = {purple,thick,->}
        ]
        \draw[draw=none] (0,0) -- (0,-.8);
        \node (label) at (-.5,2.6) {(a)};
        \def\latticeC{1.1}
        \draw[thin, gray] (-.3,-.3) rectangle ({\latticeC-0.3},{\latticeC-0.3});
        \def\anglet{15} 
        \foreach \x in {0,1,...,\countx} {
            \foreach \y in {0,1,...,\county} {
                \node[sitea] (a\x\y) at ({\latticeC*\x}, {\latticeC*\y}) {};
                \node[siteb] (b\x\y) at ({\latticeC*\x}, {\latticeC*\y}) {};
                \foreach \n in {0,1,...,3} {
                    \draw[orba] 
                    ({\latticeC*\x}, {\latticeC*\y}) .. 
                        controls +({(6*\n-0.5)*\anglet}:.5) and +({(6*\n+3+0.5)*\anglet}:.5) ..
                    ({\latticeC*\x}, {\latticeC*\y});
                    \draw[orbb] 
                    ({\latticeC*\x}, {\latticeC*\y}) .. 
                        controls +({(6*\n-3-0.5)*\anglet+1.5}:.5) and +({(6*\n+0.5)*\anglet}:.5) ..
                    ({\latticeC*\x}, {\latticeC*\y});
                }
            }
        }
        \draw[hopone] (a11.{\anglet}) .. controls +({\anglet}:.1) and +({11*\anglet}:.1) .. 
        (b21.{11*\anglet});
        \draw[hopone] (b11.{-\anglet}) ..
        controls +({-\anglet}:.1) and +({13*\anglet}:.1) ..  
        (a21.{13*\anglet}) node[pos=0.6,below =.0em] {\(t_1\)};
        \draw[hoptwo] (b11.{4*\anglet}) .. 
            controls +({4*\anglet}:.5) and +({14*\anglet}:.5) .. 
            (a22.{14*\anglet})
            node[pos=1,above left = -0.2em and 0em] {\(t_2\)};
        \draw[hoptwo] (a11.{2*\anglet}) .. 
            controls +({2*\anglet}:.5) and +({16*\anglet}:.5) .. 
            (b22.{16*\anglet}) 
            node[pos=1,below right=0 and -0.1] {\(-t_2\)};
    \end{tikzpicture}
    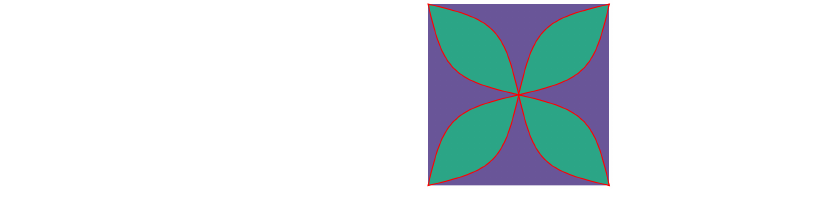
    \caption{
        Lattice model corresponding to Eq.~\eqref{eq:model-d4}.
        (a) In a square lattice, two orbitals (blue, orange) are located at the same position in a unit cell. They are fourfold rotation symmetric and each other's mirror image.
        The full set of hoppings is obtained from the drawn arrows by rotations about \(\pi/2\); per such rotation, all hoppings gain an additional minus sign relative to the ones shown in the figure. 
        Under \PT{} symmetry, both \(t_1\) and \(t_2\) are real; \(t_2\) constitutes anti-Hermitian NNN hopping.
        (b)-(e) Eigenvalue gap \(\Delta E^2\) for parameters \(t_1/t_2=100\) (\(2\), \(1\), \(0.01\)), shown globally, over the entire Brillouin zone. 
        Green (blue) indicates regions with imaginary (real) eigenvalues, bounded by exceptional lines drawn in red. 
        }
    \label{fig:model-d4}
\end{figure*}
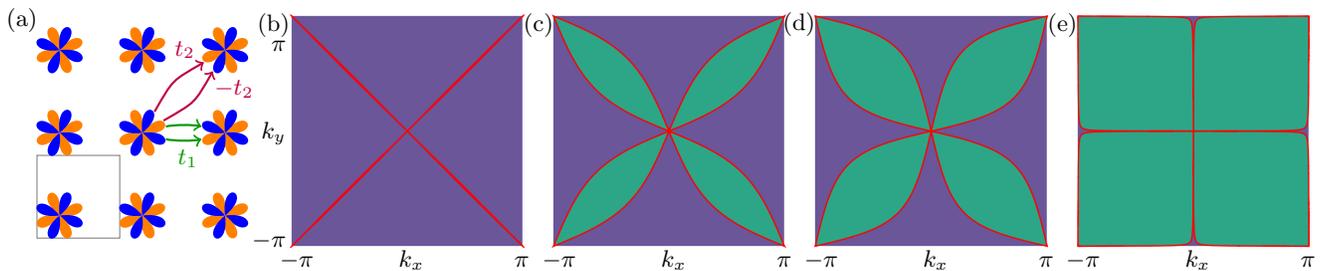

We investigate next how non-Hermitian terms affect the dispersion relation around the degenerate points of high symmetry.
Without loss of generality, we consider traceless operators.
Note that the omitted trace terms must also be symmetric under the space group action.
In particular, they generally cannot tilt the occurring dispersion.

In a neighborhood \(\bk\approx \bk^\star\), we consider the symmetry constraint~\eqref{eq:symmetry-constraint-around} order by order in the Taylor expansion of the Bloch Hamiltonian.
We find
\begin{subequations}
\label{eqs:results-spatial-only}
\begin{align}
    H^{(D_3)}(\bk) &= A (k_x \sigma_x + k_y\sigma_y) + \OO(\bk^2)
    \label{eq:results-spatial-only_D3}
    \\
    H^{(D_4)}(\bk) &= B_1 (k_x^2 - k_y^2) \sigma_x + i B_2 k_xk_y \sigma_y + \OO(\bk^4)
    \label{eq:results-spatial-only_D4}
    \\
    H^{(D_6)}(\bk) &= C\left[ (k_x^2 - k_y^2) \sigma_x \mp 2 k_xk_y\sigma_y\right] + \OO(\bk^4)
    \label{eq:results-spatial-only_D6}
    \\
    H^{(D_2^\star)}(\bk) 
        &= i E_1 k_y \sigma_y + E_2 k_x \sigma_z + \OO(\bk^2).
    \label{eq:space-group-d2-only}
\end{align}
\end{subequations}
\noindent
with arbitrary constants \(A,B_1,B_2,C,E_1,E_2 \in\C\), and the coordinate system chosen s.t. the \(x\)-axis constitutes a reflection axis.
The two signs correspond to the \(h=1,2\) representations of \(D_6\). 
The dispersion imposed by \(D_3\)-symmetry is precisely the low-energy excitation Hamiltonian of graphene at the \(K\) point.
Notably, under both \(D_4\) and \(D_6\) contributions at first order in \(\bk\) are forbidden, imposing a \(\bk^2\)-dispersion on the low-energy excitations.
For \(D_4\) and \(D_2^\star\) symmetries there is a second complex degree of freedom, allowing for a range of possible spectra.

Given that both spatial and non-spatial symmetries are known to constrain spectra significantly, it is quite natural to investigate their interplay.

First, Hermitian ``symmetry'', \(H=H^\dagger\), imposes real multiplicative prefactors in Eqs.~\eqref{eqs:results-spatial-only}.
The \(D_3\) and \(D_6\) low-energy excitation Hamiltonians remain qualitatively the same,  while for \(D_4\), \(D_2^\star\) symmetries, \(B_1, i B_2, iE_1, E_2 \in\R\), poses a qualitative constraint.

We next investigate \PT{} symmetry, which is of particular experimental relevance in non-Hermitian systems.
It corresponds to, e.g., balanced gain and loss in optical setups, is realized among others in photonic crystals, and reduces the codimension of nodal structures further, leading to the generic appearance of nodal lines in the two-dimensional systems considered here \cite{yangHomotopySymmetryNonHermitian2023}. 
In its most straightforward representation, it requires the Bloch Hamiltonian to be real at each \(\bk\).
For two-band Hamiltonians, this only allows for terms \(\sigma_x, i\sigma_y, \sigma_z\) with real prefactors, which fundamentally alters the results given in Eqs.~\eqref{eqs:results-spatial-only}.
For the four groups of relevance, the allowed terms at lowest order are 
\begin{subequations}
\begin{align}
    H^{(D_3,\PT)}(\bk) 
        &= A \left(k_x^2 k_y-\frac13k_y^3\right) \sigma_z + \OO(\bk^5),
    \\
    H^{(D_4,\PT)}(\bk) 
        &= B_1 (k_x^2 - k_y^2) \sigma_x 
        \label{eq:d4-pt:low-energy-expansion}
        \\\notag &\quad + i B_2 k_xk_y \sigma_y + \OO(\bk^4),
    \\
    H^{(D_6,\PT)}(\bk)
        &= C \left(k_x^3 k_y^3 -\frac3{10}(k_x k_y^5+k_y k_x^5)\right)\sigma_z + \OO(\bk^8),
    \\
    H^{(D_2^\star,\PT)}(\bk)
        &= i E_1 k_y \sigma_y + E_2 k_x \sigma_z + \OO(\bk^2).
\end{align}
\label{eqs:pt-models}
\end{subequations}
with real constants \(A,B_1, B_2, C, E_1, E_2 \in\R\), illustrated in Fig.~\ref{fig:main-result}.
The lowest-order contribution for \(D_3,\PT\) (\(D_6,\PT\)) symmetries occurs at third (sixth) order as opposed to first (second) order without \PT{} symmetry.
Under \PT{} symmetry the  \(D_4\)-symmetric (\(D_2^\star\)-symmetric) terms in Eq.~\eqref{eq:results-spatial-only_D4} (Eq.~\eqref{eq:space-group-d2-only}) are allowed for real coefficients. 
This constrains the models to an interesting corner of parameter space: the spectrum necessarily features four (two) exceptional lines that emanate from the nodal point at \(\bk^\star\).
These lie at angles \(2 \theta^{D_4} = \pm \operatorname{arctan}(2 B_1/B_2 )+\Z\pi\) and \(\theta^{D_2^\star} = \pm \operatorname{arctan}(E_2/E_1 )\) relative to the \(k_x\)-axis.

Finally, we note that time reversal \T{} acts similarly to \PT{} symmetry. 
In its simplest representation, it also conjugates the Bloch Hamiltonian, with the added complication that it also inverts momentum \(\bk\mapsto-\bk\).
Models with point groups that contain a half turn are \(\mathcal{P}\)-invariant, and hence, the two symmetries are equivalent.
In fact, we find that for fixed points under inversion, the two constraints are equivalent for all representations discussed. 
We expect a further interesting constraint for the \(K\)-point in p6mm, which is mapped to \(K'\) by both \T{} and a rotation about \(2\pi/6\); the specifics depend on the given representation of the rotation.
We conclude by noting that this means \T-invariance also leads to lines of EPs crossing at \(\mathbf{k}=\bk^\star\) for \(D_4\) and \(D_2^\star\) symmetries, as shown for \PT{} in Fig.~\ref{fig:main-result}.

We realize the allowed first-order terms
Eqs.~\eqref{eq:results-spatial-only_D4}~and~\eqref{eq:space-group-d2-only} in the overall \(D_4\)-symmetric, respectively, \(D_2^\star\)-symmetric, models,
\begin{align}
    H_4(\bk)=&
        2 t_1 [\cos(k_x) - \cos(k_y)]~\sigma_x 
    \label{eq:model-d4}
    \\\notag
        &+ 2i t_2 [\cos(k_x-k_y)-\cos(k_x+k_y)]~\sigma_y
    \\
    H_2(\bk)=&
        2  i \tilde{t}_1 \sin(k_y)~\sigma_y + 2 \tilde{t}_2 \sin(k_x)~\sigma_z 
    \label{eq:model-d2}
\end{align}
where we neglected the traceful parts \(d_0(\bk) I_2\) as we are mainly interested in the gap structure.

We show the spectrum of \(H_4\) in Fig.~\ref{fig:model-d4} under \PT{} symmetry, where both models show exceptional lines spanning the Brillouin zone and intersect at the high-symmetry points.
At these points, we recover low-energy expansion~\eqref{eq:d4-pt:low-energy-expansion} with \(B_1 =t_1\) and \(B_2 = 4 t_2 \).
Note that the only other allowed nearest-neighbor hopping terms contribute proportional to \(\sigma_0\) and do not affect the gap structure, since terms within the same sublattice of the form \(f(\bk)\sigma_z\) must satisfy \(f(k_x,k_y)=-f(-k_x,k_y)=-f(k_y,k_x)\) and thus vanish on the reflection axes.

In this work, we surveyed the possible nodal structures constrained by the two-dimensional wallpaper groups.
We found seven space groups with high-symmetry momenta at which the crystalline symmetry alone enforces nodal points.
The surrounding dispersion is particularly interesting in distinction to known Hermitian results when, on top of its crystalline structure, the system is \T{} or \PT-symmetric.
As our results are derived from symmetry considerations, we expect them to be valid also outside of the tight-binding approximation and relevant to experimental systems in several realms of physics, such as photonic crystals, which can be constructed corresponding to a given spatial symmetry\cite{wangEffectsShapesOrientations2001,luSymmetryprotectedTopologicalPhotonic2016}. 
The physics of non-Hermitian systems with open boundaries generally differs qualitatively due to the non-Hermitian skin effect\cite{yaoEdgeStatesTopological2018,leeAnomalousEdgeState2016,xiongWhyDoesBulk2018,kunstBiorthogonalBulkBoundaryCorrespondence2018,okumaNonHermitianTopologicalPhenomena2023,linTopologicalNonHermitianSkin2023,okumaTopologicalOriginNonHermitian2020,jiangReciprocatingBipolarNonHermitian2023,zhangReviewNonHermitianSkin2022}. However, the symmetries considered here inhibit skin effects, and spectra of lattice models, such as Eqs.~\eqref{eq:model-d4}~and~\eqref{eq:model-d2}, do not change significantly upon changing boundary conditions, as may be expected from the Bloch spectra being confined to the real and imaginary axis. This corroborates the physical relevance of our results.

While the simple representations used here are relevant to photonic and metamaterial setups, we expect different representations of \PT{} or \T{} symmetry, or the full representation theory of magnetic space groups, to produce further new and non-Hermitian nodal structure, as the extended symmetry may mix different irreducible representations of the space group \cite{yangHomotopySymmetryNonHermitian2023}.
Another exciting prospect of this consideration is the emergence of higher order degeneracy. 
A hint that this can be fruitful is that $\PT$-symmetry alone makes third order EPs generic and stable in two dimensions \cite{mandalSymmetryHigherOrderExceptional2021,delplaceSymmetryProtectedMultifoldExceptional2021}. 

An even more ambitious outlook is to extend the ana\-lysis given here to three dimensions. 
This is a major task given that there are 230 space groups in three dimensions, and that three-dimensional non-Hermitian systems generically feature exceptional lines even in the absence of symmetry. This would amount to a major step toward a photonics counterpart to the paradigm of topological quantum chemistry \cite{bradlynTopologicalQuantumChemistry2017}.

\begin{acknowledgments}
JLKK and EJB were supported by the Swedish Research Council (VR, grant 2018-00313), the Wallenberg Academy Fellows program (2018.0460) and the project Dynamic Quantum Matter (2019.0068) of the Knut and Alice Wallenberg Foundation, as well as the Göran Gustafsson Foundation for Research in Natural Sciences and Medicine.
\end{acknowledgments}

\section*{Author Declarations}

\subsection*{Conflict of Interest}
The authors have no conflicts to disclose.

\subsection*{Author Contributions}
\textbf{JLKK}: 
    supervision (supporting);
    validation (lead);
    formal analysis (equal);
    visualization (lead);
    writing -- original draft preparation (lead);
    writing -- review and editing (equal).
\textbf{FH}: 
    Formal analysis (equal);
    writing -- review and editing (equal).
\textbf{EJB}: 
    conceptualization (lead); 
    funding acquisition (lead);
    supervision (lead);
    writing -- review and editing (equal).

\subsection*{Data Availability}
The data that support the findings of this study are available within the article.

\section*{References}
\bibliography{main}

\end{document}